\documentclass[a4paper,11pt]{article}

\usepackage{amsmath}

\usepackage{graphicx}

\usepackage{epsfig}

\usepackage{here}

\newcommand{\beq}{\begin{equation}}

\newcommand{\eeq}{\end{equation}}

\newcommand{\beqa}{\begin{eqnarray}}

\newcommand{\eqa}{\end{eqnarray}}

\title{The time evaluation of resistance probability of a closed community against
occupation in a Sznajd-like model with synchronous updating: A
numerical study}

\author{\centerline{Ekrem Ayd\i ner}\\
\textit{Department of Physics, Faculty of Arts and Sciences}\\
\textit{University of Dokuz Eyl\"{u}l, Izmir, Turkey} \\
\textit{E-mail: ekrem.aydiner@deu.edu.tr}}

\begin{document}

\maketitle

\section*{Abstract}

In the present paper, we have briefly reviewed Sznajd's
sociophysics model and its variants, and also we have proposed a
simple Sznajd like sociophysics model based on Ising spin system
in order to explain the time evaluation of resistance probability
of a closed community against occupation. Using a numerical
method, we have shown that time evaluation of resistance
probability of community has a non-exponential character which
decays as stretched exponential independent of the number of
soldiers in a one-dimensional model. Furthermore, it has been
astonishingly found that our simple sociophysics model belongs to
the same universality class with random walk process on the
trapping space.

\vspace*{25pt} \noindent {\bf Keywords:} Ising Model; Politics;
Random Walk; Sociophysics; Sznajd Model.

\newpage

\section{Introduction} %\textbf{Introduction}
Modelling of the some social phenomenon using the Ising spin
system has been of considerable interest in the statistical
physics. One of the important study in this area is undoubtedly
the Sznajd model \cite{1} which deals with opinion evolution in
closed community. In the Sznajd model, each site of a one- or
two-dimensional lattice carries a spin that can be either up
(Republican) or down (Democrat) and represent one of two
possibilities on any questions. Two neighboring parallel spins
i.e., two neighboring people sharing the same opinion, convince
their neighbors of this opinion. If they do not have the same
opinion, then either they do not influence their neighbors or they
bring their neighbors to the opposite position. In this model, the
system evolves from one time step to another through a random
sequential updating mechanism and it always reaches an overall
consensus, i.e., the system ends up in a fixed point: either all
spins point up or all point down.

Many variants of this Sznajd model are studied. Various rules are
defined and discussed in \cite{2}. In all of them, spins pairs are
selected randomly for trying to convince their neighbors. Stauffer
stated that most variants of the systems always end up in a fixed
point, where all spin are the same, or where the opinions are
ordered anti-ferromagnetically in one or two dimensional for these
rules \cite{3}.

Memory of the system had been discussed based on the Sznajd model
with synchronous updating in Ref.\cite{4}. In the case of a system
corresponding to a community always going to an overall consensus,
this situation causes a dictatorship (all spins parallel) in the
Sznajd model. Therefore, to avoid dictatorship one can introduce a
small number of non-conformist which are not convinced by the
Sznajd rule (Schneider \cite{5}), or the Sznajd model can be
studied in disordered lattices so that this is more realistic
since human society does not follow a square lattice with every
person having exactly four neighbors. The simplest case is a
diluted square lattice, where every site is either empty or having
a single spin \cite{5}. To simulate political election results,
the Sznajd model was generalized from two different opinion to
number $q$ of opinions \cite{6} and also Stauffer discussed Sznajd
model with number $q$ of opinions. He stated that there is
reachable consensus for $q\leq3$ but not for $q\geq4$ \cite{7}.
Furthermore, a different persuasion process with a continuous
spectrum of opinions between 0 and 1 were studied by Deffuant et
al \cite{8}.

Another generalization of Sznajd model to a triangular lattice
with spreading of mixed opinion and with the pure
antiferromagnetic  opinion was studied by Chang \cite{9}, who
found the fixed points for all values of initial concentration of
down spins for the mixed case of ferromagnetic and
antiferromagnetic opinions. Furthermore, a $d$-dimensional Sznajd
model of consensus finding process was simulated with $1\leq d
\leq4$ for the two possible opinions, and the density of never
changed opinion (which correspond to persistent spins in Ising
model \cite{10}) during the Sznajd consensus-finding process
decays with $t$ as $t^{1/\theta }$ were discussed. It was found
that exponent $\theta$ is compatible with the Ising value for one
dimension although it is not compatible for higher dimensions
\cite{11}.

Although, political cases that have two state opinion are quite
extreme example, such phenomena appears when a referendum has been
made in a community or in a country. People vote either "Yes" or
"No" in a referendum such as election either Democrats or
Republicans in US political conditions.

Consequently, it seems that the Sznajd model \cite{1} and its
variants \cite{2,3,4,5,6,7,9,10,11} suggests to be able to make
the physical model of some sociological or political events like
Galam's sociophysics models \cite{12,13}, although there were
resistance against modelling of the some sociophysics events from
physicists \cite{13}.

The intent of this study motivated from Sznajd's \cite{1},
Stauffer's \cite{14} and Galam's \cite{13} papers is to introduce
a simple sociophysics model to explain the time evaluation of
resistance probability of a closed community against occupation.

\section{The Model} \noindent
Everybody knows that human history has been filled with conflicts,
invasion or wars among countries or community. In historical age,
either some countries (or community) have taken over other
countries or some countries (or community) have been occupied by
other countries in a military or different way. It is difficult to
show that a country or a society had been able to stay out of such
a process.

Of course, conflicts, invasions, occupations and wars are quite
complicate phenomen which generally occur on the economical,
military, sociological, political, humanity levels etc. Therefore,
it is impossible to suppose a sociophysics model comprised of all
events in a conflicts, an invasion or a war.

Nevertheless, in order to explain the dynamics of some
socio-political events, a sociophysical model can be set based on
Ising spin systems just like Sznajd model even though it has to
contain some definite limitations.  In this study is considered a
process in which a community (or a country) is under military
occupation. To modelling such an occupation processes, some
assumptions, of which details are given as follows, are
considered.

\begin{figure}[htbp]%ORIGINAL SIZE:
\centerline {\epsfig{figure=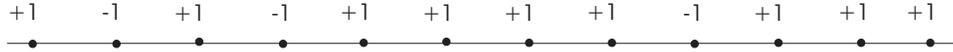,width=30pc}}
\caption{Schematic representation of occupation process. Host
people and occupying soldiers are represented with spin up (+1)
and spin down (--1) respectively. Soldiers are distributed
randomly among host people in the one-dimensional discrete
space.}\label{Fig1}
\end{figure}

Initially, let us consider one-dimensional discrete lattice. Each
site carriers a spin, $S$, which is either up or down as shown in
Fig.\ref{Fig1}. Host people of community and soldiers are
represented with spin up (+1) and spin down (-1) respectively.

Down spins are distributed randomly on the lattice governed by
volume exclusion principle as seen in Fig.\ref{Fig1}. This means
that only one person (host or soldier) occupies a site. The number
of soldiers is only a small fraction of the total population and
then the density of soldiers, $\rho$ (for one-dimensional model),
is defined as number of soldiers over total population.

We assumed that there is a over all consensus among host people in
the community against occupation even if some exceptions exist.
One expects that host people obey to this consensus at least
initially. In this sense, community behaves as polarized at zero
social temperature \cite{15} against occupation just like Ising
ferromagnet at zero temperature.

As known, every person may be influenced by others, since they
interact with each other in the community. Change of opinions of
people can be simply explained by social impact \cite{16}.

But in our model, soldiers have not been influenced by the host
people. Their opinion about justifying the occupation does not
change during the occupation process, since they may be stubborn,
stable or professional etc. In this sense, it may be assumed that
soldiers behaves like persistent spins in Ising spin system
\cite{10}, which are change their orientations corresponding to
their opinions. In addition, mission of the soldiers is not only
occupation, at the same time, they also want to convince the host
people about justification of occupation.

Unlike soldiers, it can be conjecture that host people are
influenced by soldiers even though they against occupation. They
are exposed to intensive biased information or propagation, which
play a role as persuasion mechanism.

To explain the time evolution of one-dimensional sociophysics
models, let us concentrate on how our mechanism progresses. At
this point, we shall define a resistance probability to occupation
for all people (host or soldier) at any site $i$. For any soldier,
the resistance probability, $W_i$, could be zero during an
occupation process since they believe in the justification of
occupation. On the other hand, for host people at any site $i$,
the resistance probability $W_i$ is equal to one initially since
they obey the community consensus.

In this manner, soldiers affect neighbors who are host people, in
other words, a host at any site $i$ is influenced by the
neighbors. Effected people may change their own opinions depending
on resistance probability of the nearest neighbors about
occupation. Namely, the phrase of 'winner takes all' \cite{17} is
invalid herein, they might take only a probabilistic manner. Such
a mechanism will depolarize the polarization (resistance
probability) of all host people.

For the any time step, resistance probability $W_i$ of any host
people at site $i$ is determined by probability of the nearest
neighbors at the previous time as
\begin{equation}\label{1}
W_{i}\left( N+1\right) =\frac{1}{2}\left[ W_{i-1}\left( N\right)
+W_{i+1}\left( N\right) \right]
\end{equation}
Eq.(\ref{1}) clearly indicate that past plays an important role in
this model. Also, the system evolves from one time step to another
through synchronized updating mechanism.

Total resistance probability of community will be depolarized by
soldiers even if we assumed that community has a consensus about
resistance to occupation initially. Total polarization for such a
system could be calculated over every host people for any time
step $N$ as
\begin{equation}\label{2}
P_r\left( N\right) =\frac{1}{m_{0}}
\overset{m_0}{\underset{i=1}{\sum}}W_{i}\left( N\right)
\end{equation}
where $m_{0}$ is the initial number of host people, $r$ represents
any configuration and the sum is over all sites. If Eq.(\ref{2})
is averaged over different soldier configurations, then it can be
rewritten as
\begin{equation}\label{3}
<P\left( N\right)>
=\frac{1}{R}\overset{R}{\underset{r=1}{\sum}}P_r(N)
\end{equation}
where $R$ is the number of different configurations.

\section{Numerical Results} 
\begin{figure}[htbp]%ORIGINAL SIZE:
\centerline {\epsfig{figure=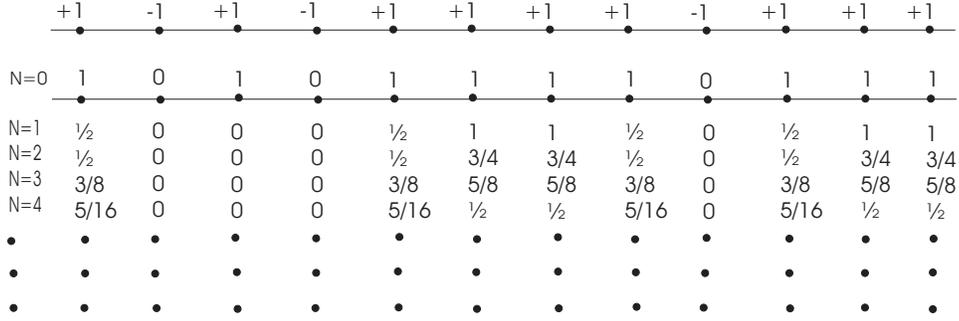,width=30pc}}
\caption{Schematic representation of the time evolution of the
resistance probability for every host people in the one
dimensional occupied community model.}\label{Fig2}
\end{figure}

\noindent
As an example, the schematic representation of the time evolution
of resistance probability due to Eq.(\ref{1}) is written using
periodic condition for every spin (host people and soldiers) is
shown in Fig.\ref{Fig2}. In this figure, the first line represents
the occupied community, and second line corresponds to its
resistance probability at time $N=0$. While probability values for
soldiers remain unchanged, it decreases for host people as seen in
Fig.\ref{Fig2}.

Evolution of resistance probabilities for each host in time $N$ is
partly illustrated in Fig.\ref{Fig2}. Fortunately, Eq.(\ref{3})
allow us to work with large lattice and over many different
configurations. For this reason, Eq.(\ref{3}) has been solved
numerically for lattice size $L=10.000$, and density of the
soldier $\rho=0.1$, number of the time step $N=5000$ and number of
the independent configuration $R=1000$ with periodic boundary
configuration by using a simple Fortran program.

\begin{figure}[htbp]%ORIGINAL SIZE:
\centerline {\epsfig{figure=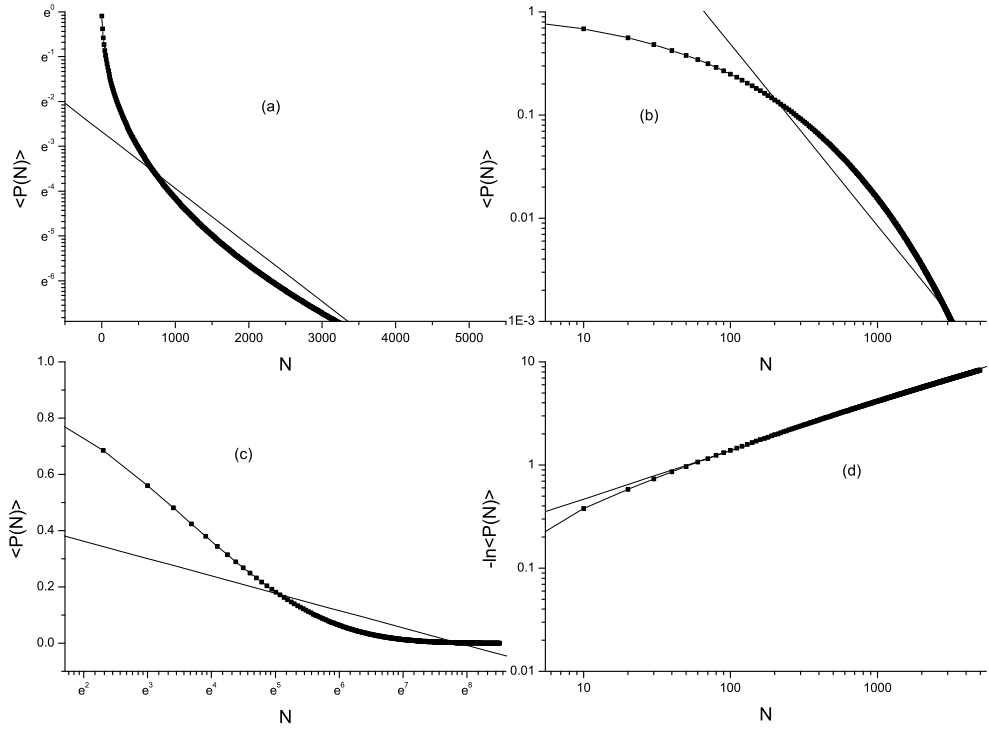,width=35pc}}
\caption{Numerical data for $\rho=0.1$ were plotted $\ln<P(N)>$
versus $N$ in Fig.3a, $\log<P(N)>$ versus $\log N$ in Fig.3b,
$<P(N)>$ versus $\ln N$ in Fig.3c, and $\log(-\ln<P(N)>)$ versus
$\log N$ in Fig.3d. Solid-dot lines indicate numerical data and
solid lines represent fitting curves in all figures.}\label{Fig3}
\end{figure}

In order to understand clearly the time evolution of
one-dimensional sociophysics model under conditions mentioned
above, the numerical data were plotted as $\ln<P(N)>$ versus $N$
in Fig.\ref{Fig3}a, as $\log<P(N)>$ versus $\log N$ in
Fig.\ref{Fig3}b, as $<P(N)>$ versus $\ln N$ in Fig.\ref{Fig3}c,
and also as $\log(-\ln<P(N>)$ versus $\log N$ in Fig.\ref{Fig3}d.

It is explicitly seen from Fig.\ref{Fig3}a, Fig.\ref{Fig3}b, and
Fig.\ref{Fig3}c that there are no exponential, power, logarithmic
laws dependence in the numerical data, respectively. But, as seen
in Fig.\ref{Fig3}d, numerical data well fit to stretched
exponential function as

\begin{equation}\label{4}
<P\left( N\right)>=Ae^{-aN^{\beta}}
\end{equation}
where $A$ normalization factor, $a$ relaxation constant, and
$\beta$ is the decay exponent of the averaged resistance
probability.

\begin{figure}[htbp]%ORIGINAL SIZE:
\centerline {\epsfig{figure=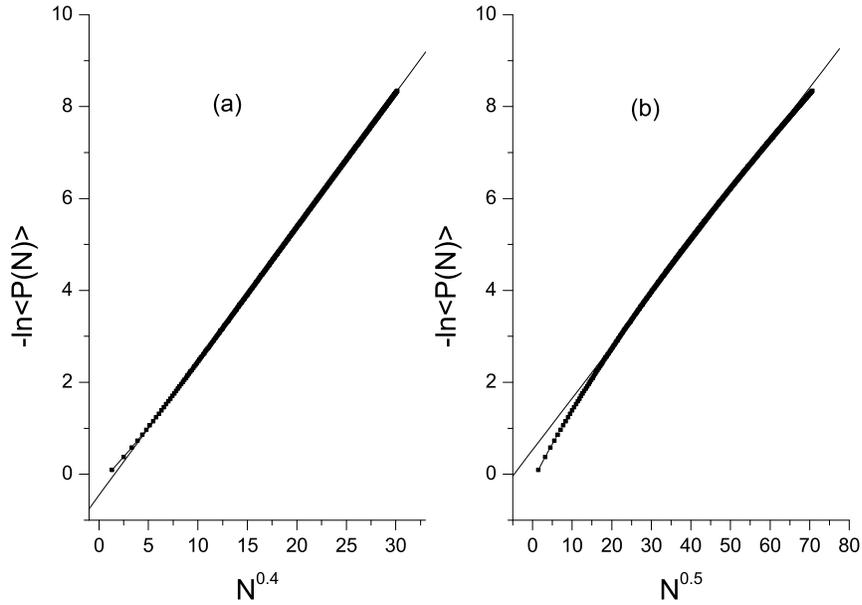,width=30pc}}
\caption{(a) $-\ln<P(N)>$ versus $N^{0.4}$ for $\rho=0.1$, and (b)
$-\ln<P(N)>$ versus $N^{0.5}$ for $\rho=0.1$. Solid-dot lines
indicate numerical data and solid lines represent fitting curves
in all figures.}\label{Fig4}
\end{figure}

In addition, it should be tested whether Fig.\ref{Fig3}d satisfies
to stretched exponential or not. If prefactor $A$ is equal to 1 in
Eq.(\ref{4}) the Fig.\ref{Fig3}d would work as stretched
exponential. However, if $A$ is less than $1$, small prefactor may
give the impression of stretched exponential form, even for
$\beta=1$. Therefore, it can be plotted $-\ln<P(N)>$ versus
suitable powers of $N$, like $N^{1/2}$, $N^{1/3}$, etc., and find
out the best straight line among the powers of $N$ for long times
(email from D. Stauffer). Hence, $-\ln<P(N)>$ was plotted versus
powers of $N$ for $\rho=0.1$ then the best straight fitting line
for long times was obtained for $\beta=0.4$ and $\beta=0.5$ as
seen in Fig.\ref{Fig4}a and Fig.\ref{Fig4}b, respectively. These
results confirm to this method used to find out stretched
exponential exponents in Fig.\ref{Fig3}d. Also, this test
indicates that prefactor in Eq.(\ref{4}) does not effect results
presented in this paper.

\begin{figure}[htbp]%ORIGINAL SIZE:
\centerline {\epsfig{figure=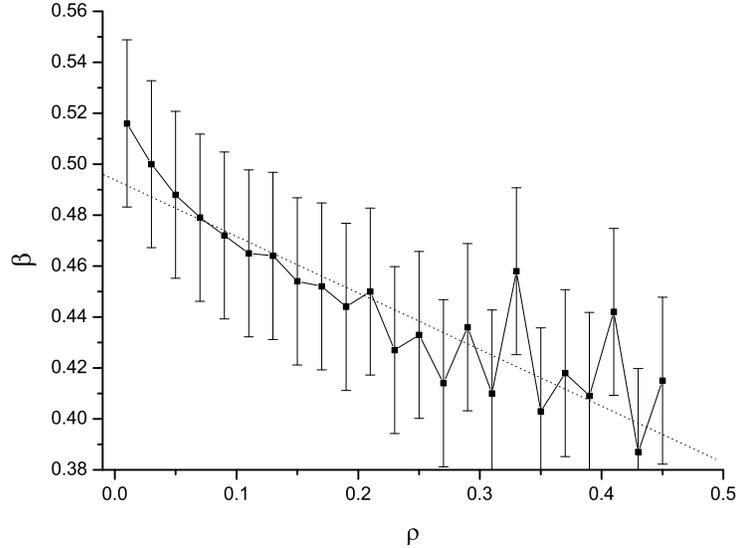,width=26pc}}
\caption{Change of the exponent $\beta$ depending on density of
soldiers $\rho$ for one-dimensional model. Solid-dot line and dot
line represent numerical data and fitting curve
respectively.}\label{Fig5}
\end{figure}

To see whether the exponent $\beta$ depends on $\rho$ or not, we
plotted $\beta$ versus $\rho$ in Fig.\ref{Fig5}. This figure shows
clearly that exponent $\beta$ is a function of density of
soldiers. As seen from Fig.\ref{Fig5}, $\beta$ almost linearly
decreases. It is inferred from Fig.\ref{Fig5} that the averaged
resistance probability $<P(N)>$ has stretched exponential
character independent of $\rho$. But, one cannot guaranty that it
is true for all $\rho$ values. As seen in Fig.\ref{Fig5}, while
$\beta$ monotonically decreases for small $\rho$ whereas
oscillating occurs in $\beta$ for large $\rho$ values, which
indicates that fatal statistical errors may occur for large $\rho$
values.

\section{Conclusions}\noindent
In all reasoning, it is inferred that numerical results obtained
from simple sociophysics model studied herein obey a stretched
exponential law independent of the number of soldiers. Also, we
find out exponent $\beta$ depend on number of soldiers. These are
quite remarkable aspects.

Stretched exponential behavior indicates mathematically that decay
for the relatively short times is fast, but for relatively long
times it is slower. One can observe that this mathematical
behavior corresponds to occupation processes in the real world. In
generally, a military occupation is realized after a hot war. The
community does not react to occupation since it occurs as a result
of defeat. People are affected easily by propaganda or other
similar ways. Therefore, it is no surprise that resistance
probability decrease rapidly at relatively short times. On the
other hand, spontaneous reaction may begin against occupation in
the community after the shock. Hence, community begins by
regaining consciousness and more organized resistance may display
difficulties for occupants. For long times, the resistance
probability decreases more slowly. This means that resistance
against occupation extends to long times in practice. At this
point, the number of soldiers is also important, because the
density of soldiers determines the speed of decaying.

Of course, the mechanism considered in this work can be regarded
as simple, but, it would be useful to understand the time
evolution of the resistance probability of the community against
to occupation in the one-dimensional model under some
considerations. We noted that more realistic sociophysics model
for better understanding the time evolution of resistance against
to occupation appearing may be devised on the randomly diluted
square lattice, where every site is either empty or carrier one
spin which represent a host people or soldier randomly. Also, in
that model both host people and soldiers move on the lattice via
diffusion. Hereafter, also the sociophysics model mentioned above
will be treated in the near future.

Finally, it is surprisingly seen that there is a connection simple
sociophysics model of which details are given in this paper with
one-dimensional random walk processes in the presence of a
periodic distribution of traps \cite{18,19,20}. In the both
models, the resistance probability for any host people and the
survival probability for any walker are given by Eq.(\ref{1}) at
any time. Since the time evolution of probabilities in these
models arise from the same result, hence, one can suggest that the
sociophysics model herein and random walk model on the trapping
lattice are likely to be in the same universality class.

\section*{Acknowledgements}
I would like to thank to Dietrich Stauffer for his valuable
corrections and contributions about this work.


\section*{References}

\begin{thebibliography}{000}

\bibitem{1}
K.Sznajd-Weron and J.Sznajd, \newblock Int. J. Mod. Phys. C, {\bf
11} (2000) 1157.

\bibitem{2}
D.Stauffer, A.O.Sousa, S.Moss de Olivera, \newblock Int. J. Mod.
Phys. C, {\bf 11} (2000) 1239.

\bibitem{3}
D.Stauffer, \newblock AIP Conf. Proc. {\bf 190} (2002) 147;
J.Art.Soc.Soc.Sim., {\bf 5}, No.1 paper 4 (2002)
(http://www.soc.surrey.ac.uk/JASSS/5/1/4.html).

\bibitem{4}
L.Sabatelli, P.Richmond, \newblock Int. J. Mod. Phys. C {\bf 14}
(2003) 1223.

\bibitem{5}
J. J. Schneider, \newblock Int. J. Mod. Phys. C {\bf 15}, issue 5
(2004); A.A.Moreira, J.S.Andrade Jr., D.Stauffer, \newblock Int.
J. Mod. Phys. C {\bf 12} (2001) 39.

\bibitem{6}
A.T.Bernardes, U.M.S.Costa, A.D.Araujo, D.Stauffer, \newblock Int.
J. Mod. Phys. C, {\bf 12} (2001) 159.

\bibitem{7}
D.Stauffer, \newblock Int. J. Mod. Phys. C {\bf 13} (2002) 315.

\bibitem{8}
G.Deffuant, D.Neau, F.Amblard, G.Weisbuch, \newblock Adv. Complex
Syst. {\bf 3} (2000) 87.

\bibitem{9}
I.Chang, \newblock Int. J. Mod. Phys. C {\bf 12} (2001) 1509.

\bibitem{10}
B.Derrida, A.J.Bray, C.Godreche, \newblock J. Phys. A {\bf 27}
(1994) 357.

\bibitem{11}
D.Stauffer, P.M.C.Oliveira, \newblock Eur. Phys. J. B {\bf 30}
(2002) 587.

\bibitem{12}
S.Galam, \newblock Physica A {\bf 333} (2004) 453.

\bibitem{13}
S.Galam,  \newblock Physica A {\bf 336} (2004) 49, and the
references therein,

\bibitem{14}
D.Stauffer,  \newblock Physica A {\bf 336} (2004) 1.

\bibitem{15}
F.Schweitzer, J.A.Holyst, \newblock Eur. Phys. J. B {\bf 15}
(2000) 723.

\bibitem{16}
B.Latan\'{e}, \newblock Am. Psychologist {\bf 36} (1981) 343.

\bibitem{17}
D.Stauffer and J.S.Sa Martins,  \newblock Physica A {\bf 334}
(2004) 558.

\bibitem{18}
R.Cohen, K.Erez, D.ben-Avraham and S.Havlin, \newblock Phys. Rev.
Lett., {\bf 85} (2000) 4626.

\bibitem{19}
D.S.Callaway, M.E.J.Newman, S.H.Strogatz and D.J.Watts,
\newblock Phys. Rev. Lett., {\bf 85} (2000) 5468.

\bibitem{20}
S.Havlin, G.H.Weiss, J.E.Kiefer and M.Dishon, \newblock J. Phys.
A: Math. Gen., {\bf 17} (1984) L347.

\end{thebibliography}
\end{document}